\documentclass[10pt,a4paper]{article}

\usepackage[ruled]{algorithm2e}
\usepackage{amssymb}
\usepackage{amsmath}
\usepackage{array}
\usepackage{mdwmath}
\usepackage{mdwtab}
\usepackage{url}
\usepackage{graphicx}
\usepackage[normalem]{ulem}
\usepackage{paralist}
\usepackage{placeins}

\begin{document}

\title{Data-Driven Application Maintenance: Views from the Trenches}
\author{
Janardan Misra, Shubhashis Sengupta \\ Milind Savagaonkar, Divya Rawat, Sanjay Podder \and
Accenture Labs, Bangalore, India\and
\{ {\sf janardan.misra,shubhashis.sengupta,} \\ {\sf milind.savagaonkar,d.b.rawat},\\ {\sf sanjay.podder}\}{\sf @accenture.com}
}

%\author{
%Janardan Misra \\
%Accenture Labs, Bangalore, India\\
%janardan.misra@accenture.com
%\and
%Shubhashis Sengupta\\
%Accenture Labs, Bangalore, India\\
%shubhashis.sengupta@accenture.com
%\and
%Milind Savagaonkar\\
%Accenture Labs, Bangalore, India\\
%milind.savagaonkar@accenture.com
%\and
%Divya Rawat\\
%Accenture Labs, Bangalore, India\\
%d.b.rawat@accenture.com
%\and
%Sanjay Podder \\
%Accenture Labs, Bangalore, India\\
%sanjay.podder@accenture.com
%}

\maketitle

\begin{abstract} In this paper we present our experience during design, development, and pilot deployments of a data-driven machine learning based application maintenance solution. We implemented a proof of concept to address a spectrum of interrelated problems encountered in application maintenance projects including duplicate incident ticket identification, assignee recommendation, theme mining, and mapping of incidents to business processes. In the context of IT services, these problems are frequently encountered, yet there is a gap in bringing automation and optimization. Despite long-standing research around mining and analysis of software repositories, such research outputs are not adopted well in practice due to the constraints these solutions impose on the users. We discuss need for designing pragmatic solutions with low barriers to adoption and addressing right level of complexity of problems with respect to underlying business constraints and nature of data.
\end{abstract}
%\keywords{application maintenance, bug deduplication, assignee recommendation, theme mining, business process mapping, text analysis, machine learning, NLP}   

\section{Introduction}

Ticket or Incident Management is an important ITIL~\cite{Ref1} process, where an incoming ticket logged by the users of the system is analyzed and appropriate measures are taken for remediation within pre-specified time constraints. In terms of an activity pipeline, the tickets are first triaged, starting with a check for duplication so that one does not start working on a ticket which has already been closed or remediated, the tickets are then clustered based on themes around business process or techno-functional areas~\cite{Ref2}; further, the tickets are assigned to service engineers who have already worked on similar incidents or technology; the service engineers may take recourse to knowledge management tools which may help them to resolve the problem in a guided way by providing contextual search facility on product knowledge base or literature~\cite{Ref3}. 

Volume of research has been conducted on each of the steps of the process. The problem of triaging and identifying duplicate incidents or tickets has been discussed in ~\cite{Ref4}, especially in the context of mining large-scale open source and other software repositories and developer comments. Mapping to business processes has been studied in~\cite{Ref6}. Recommendation of assignee has been discussed in~\cite{Ref7}. % usage of knowledge management for search and resolution support has been investigated in [Ref8]. 

While there is no dearth in investigation on and application of advanced techniques like Machine Learning (classification and clustering), Natural Language Processing and Guided Search; in practice, the problem remains largely unsolved. There are several contributing factors that make ``in the trench'' application of sophisticated technologies challenging. We list a few of them –
\begin{description}
\item[The data conundrum] Industry data is either very noisy or very sparse. This is especially true when the systems have wide variety of applications and usage scenarios. This makes data sampling and preparation for machine learning fairly difficult.
\item[SME support] Most classification techniques would require proper identification of features which are discriminating enough to render the data set classifiable. Often times, we find that these discriminating features are buried in the semantics of the description and it becomes very difficult for a data-scientist to conduct proper feature engineering without help from the subject matter experts (SMEs) to understand the semantics. However, the SMEs may often not have enough time to support.
\item[Cost and Time to market] In practice, for the solution to be usable, the time to maturity would have to be short and implementable without much penalty in terms of cost and time. This is because such systems are often used to augment the support engineers, who may not have enough expertise and training to deal with intricacies.
\end{description}

In this report, we describe a ``shop usable'' for some aspects of the incident management problem and share experience from the implementation details. We believe that this paper would illustrate the contours and challenges of a practical research where unsupervised semantic similarity and ML techniques have been deployed to provide good enough solution. We also share some thoughts on the challenges for roll-out and feedback from the actual users of this incident management system.

{\sf Notations:} In the rest of the paper, we will use `ticket', `incident', and `bug' interchangeably. The PoC prototype tool will be referred henceforth as \textsl{CogAM} (Cognitive Application Maintenance).

\section{Problem Scenarios}\label{prob_scene}

\begin{figure}
	\centering
		\includegraphics[scale=0.5]{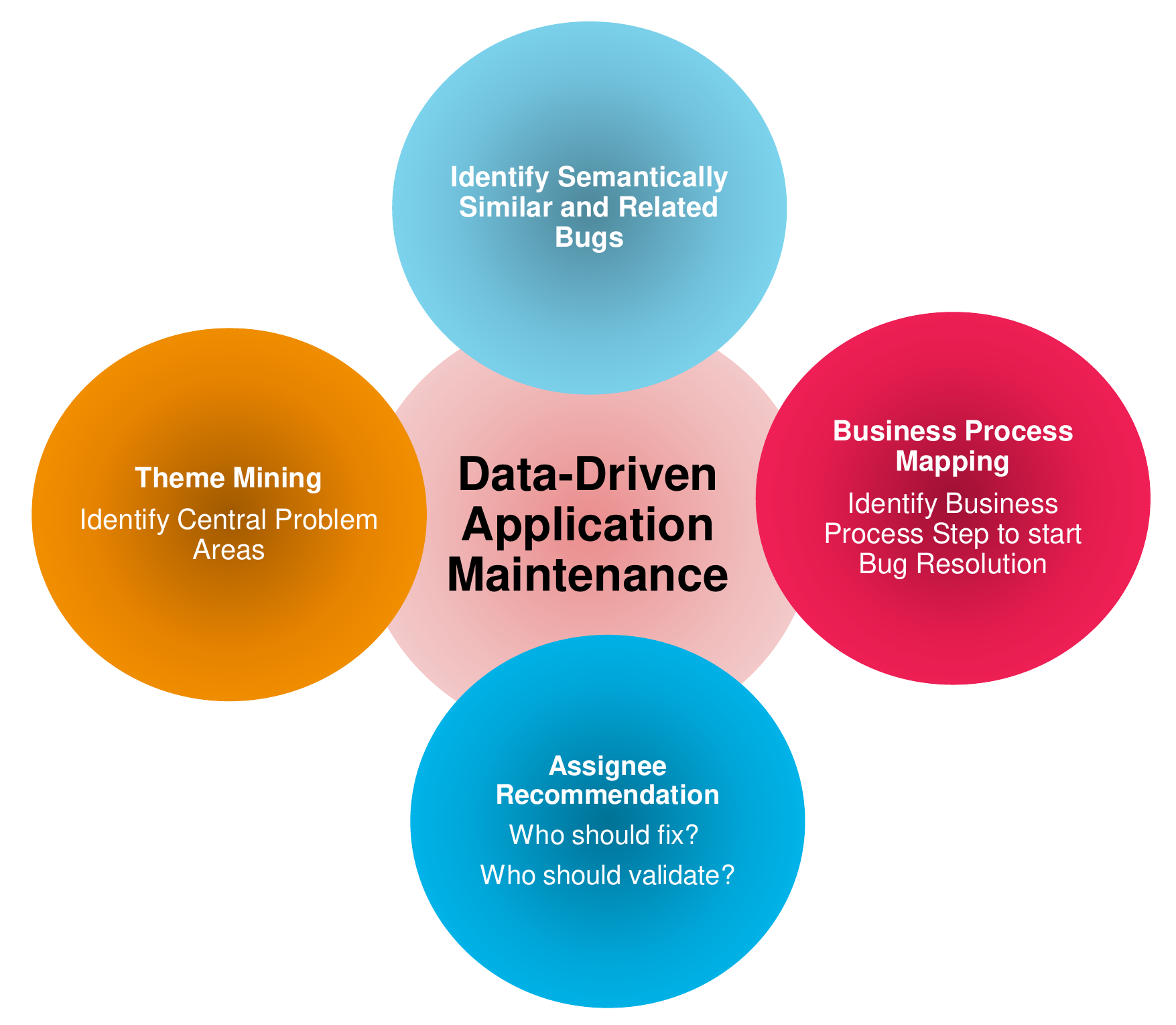}
		\caption{Problem Scenarios}\label{pv}
\end{figure}

\subsection{Bug Deduplication: Identify Similar Incidents}

\sf{Current Scenario:} When a new ticket is logged, users need to search existing ticket logs to find out if a duplicate bug has been already resolved/closed in past. Effectiveness of the search process not only depends upon nature of search query in relation to existing ticket log data but also on knowledge and expertize of the user. Even though existing application maintenance (AM) tools (e.g., Remedy~cite{remedy}, ServiceNow~\cite{servicenow}) offer many features to manage incidents and other associated processes, they fall short when it comes to offering plain text based semantic relevance search (often found in the realm of search engines). This makes it hard for users having relatively less experience to determine if semantically similar defects have already been raised in past and what actions were performed for those bugs.  

{\sf Solution Idea:} In order to keep barriers to adoption low, we adopted IR (information retrieval) based solution (similar to ~\cite{amoui2013search}) by aiming to enable search based bug deduplication and identification of semantically related incidents. IR based approaches are largely unsupervised in nature, hence saving initial efforts from SMEs towards training the tool and appear to solve deduplication problem in AM domain at right level of complexity. It's success is based upon implicit assumption that if two bugs are similar (or related), they would have overlapping surface form representations (ref. `distributional hypothesis'~\cite{}).

Solution starts with generating a temporal informational model of incidents and their associations with assignees, business processes, and other semantically relevant factors. This model is used to estimate strengths of semantic associations to recommend related incidents (including duplicates), suggest assignees, identify potential business process associations, and pointers to potential resolutions. Ability to learn incrementally is designed to improve relevance of results from user interactions. Performance of CogAM could be improved further if it is coupled with domain knowledge sources, e.g., alias glossary consisting of synonymous terms.

\subsection{Mapping Tickets to Business Processes}

\sf{Current Scenario:} When a new ticket is logged in an AM tool, SME may map reported incident to specific business process as a starting point to provide right level of intervention. Correctness of mapping primarily depends upon knowledge and expertize of the SME in conjunction with the degree of comprehensiveness of business process documentation. 

{\sf Solution Idea:} Using existing log of tickets and associated business process mappings as recoded in the ticket log, CogAM would learn semantic associations between tickets and business processes by way of generating a ML model. When a new ticket arrives, CogAM would identify most probable business process, which current ticket should be mapped to and would recommend that to user. %in ranked order together with degree of confidence measure and associated explanatory analysis. 
Based upon user acceptance or rejection of the recommendation, CogAM incrementally learns and improves its performance on the ticket mapping process.

\subsection{Assignee Recommendation}

\sf{Current Scenario:} When a new ticket is logged, based upon expertize, availability, and other factors, suitable team member is selected and ticket is assigned to him/her for resolution. If an AM tool has to be used for an automated assignment, it requires manual configuration of parameters and tickets in the AM tool. However, state-of-the-art AM tools ignore historical evidence from past experience of working on similar tickets and are generally designed as rule based systems. 

however, current practice is far from being optimal. From discussions with SMEs it turned out that when team sizes are not small (> 20) and ticket influx rates are high (> 50 per week), not always first time fixes of tickets were found to be correct and for 12-15\% occasions tickets were either reopened owing to incorrect fixes or got reassigned to different team member.   

{\sf Solution Idea:} When a new ticket arrives, CogAM parses its details and would identify potential relationship with currently available team members (having worked in past) to map the ticket to those team members who could potentially be assigned to it because they worked on similar tickets in past. CogAM would produce a ranked list of team members so that right decision could be made even when the most suitable team member is not available. Based upon user acceptance or rejection of the recommendation, CogAM would incrementally learn and improve its performance on the assignment process. Performance of CogAM could be improved further if it is coupled with domain heuristics (e.g., documented expertize of the team members). 

\subsection{Theme Mining: Identifying Central Problem Areas}

\sf{Current Scenario:} When a ticket is raised, it may sometimes get tagged against high level functionality or component of the system or type of error potentially underlying the bug. Beyond this tagging, getting any further insights on the problem areas around which bugs are being raised remains limited to subjective experiences of the SMEs. 

Large scale collective analysis of ticket text is necessary to identify central problem areas around which bugs are being raised over a course of time and semantic relationships among these problem areas. 

Such insights enable better decision making towards allocating resources during design of newer versions of the applications and also during their maintenance.        

{\sf Solution Idea:} Using existing log of tickets CogAM applies combination of theme mining techniques including topic modelling, frequency based centrality, latent semantic analysis based centrality~\cite{raise2016} on the terms (noun phrases and verb phrases) extracted from the bug data. Based upon that it would identify central terms with at least 85\%  bug coverage and their spread within user defined tagging (e.g., within modules).

\section{Tool Design}\label{design} 

\begin{figure*}[ht!]
	\centering
		\includegraphics[width=\textwidth]{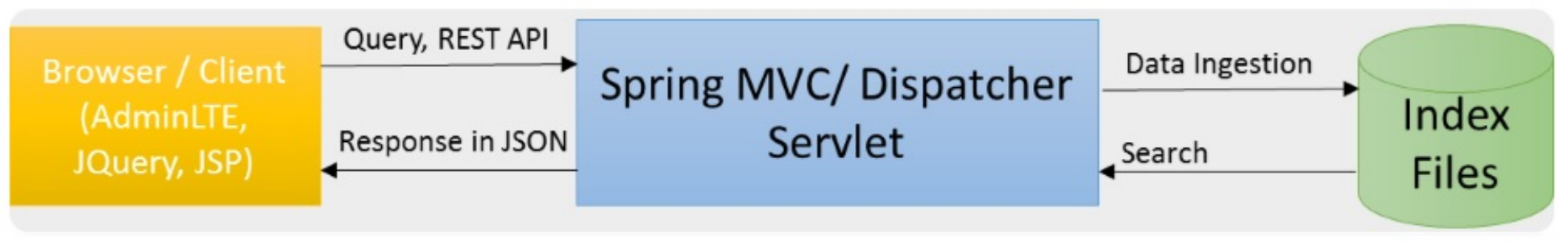}
		\caption{High level architecture}\label{hla}
\end{figure*}

\begin{figure*}
	\centering
		\includegraphics[width=\textwidth]{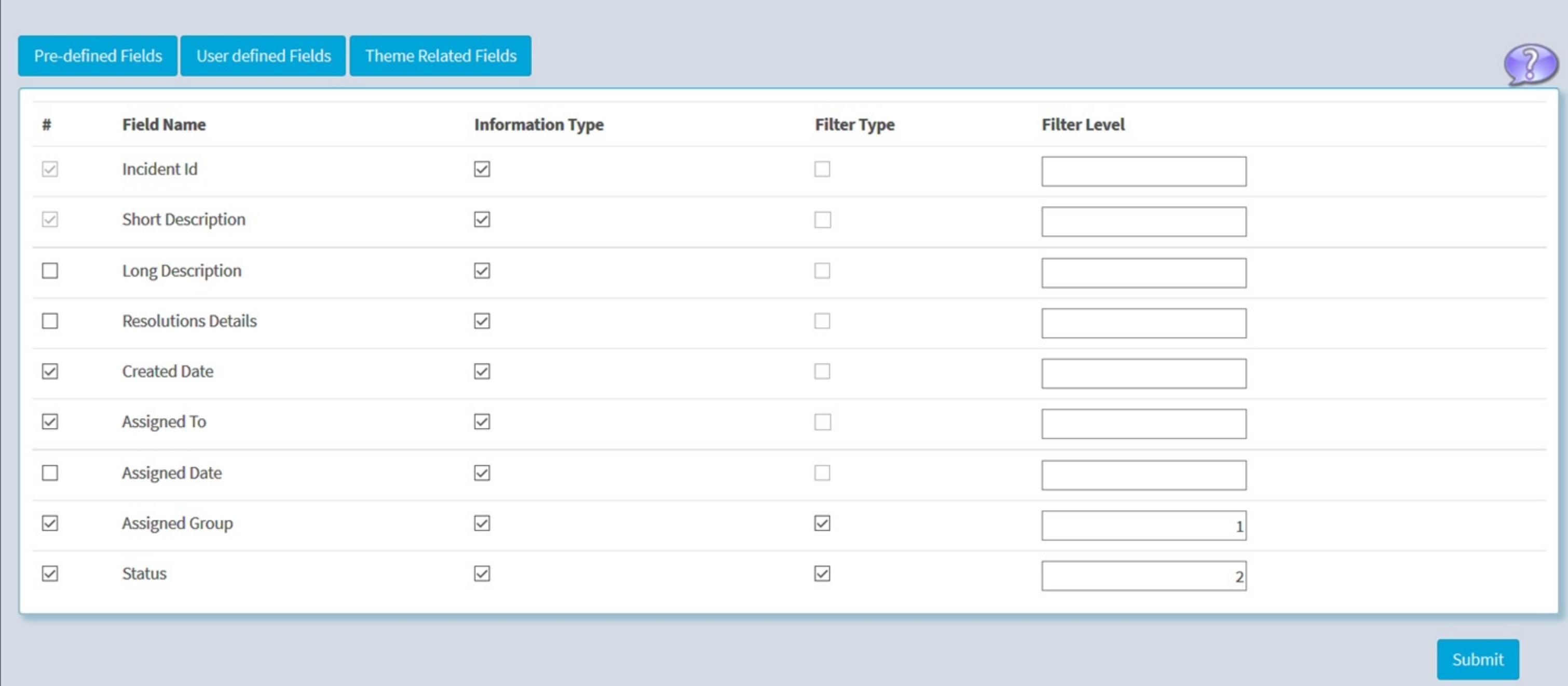}
		\caption{Field configurations}\label{config}
\end{figure*}

CogAM is implemented as web application in Java using Spring MVC architecture~\cite{mvc}. User experience is the most important thing in any application. We used AdminLTE template~\cite{lte} that is built on Bootstrap 3, JSP and JQuery to make user interface (UI) more user friendly, responsive and informative. We use Apache Lucene~\cite{lucene} for indexing and retrieving data and OpenNLP~\cite{8} for natural language processing. The high level architecture is shown in Figure~\ref{hla}.
% \FloatBarrier
The first step is to configure fields that are used to retrieve data and apply filters to refine search results. Tool provides some fields as pre-defined fields like Incident Id, Short description, Long Description etc. Beyond these, tool allows users to define custom fields as per business requirements. Figure~\ref{config} describes filed configuration. 

A field can be marked as information type where details of that particular field will be available in output screen or filter type with filter level where the field can be used as filter during search to refine results as per the requirements of the user. Filter level is used to define filter hierarchy e.g. Country, State, City is filter hierarchy while providing address details.

The second step is to ingest data. The tool supports MS Excel format for input data. Table~\ref{t1} presents input data model, in which inputs to the designed tool need to be given as inputs.
\begin{table}
\centering
\label{t1}
\begin{tabular}{l|l}
\hline
\hline
{\bf Field Name} & {\bf Field Description} \\
\hline
\hline
{ID} & Unique identifier of a ticket \\ \hline
{Summary} &  Brief description of error \\ \hline
{Description} &  Detailed description of error\\ \hline
{Assignee} &  Team member who fixed the bug \\ \hline
{Business Process} &  Associated business process step \\\hline
{Created Date} &  Date and time of creation of ticket \\\hline
{Module} &  Components or functionality affected \\\hline
{Priority} &  Criticality level assigned to the ticket \\ \hline
{Status} &  [Closed | Open | Assigned |  \\ 
&            Re-Opened | ...] \\ \hline
\end{tabular}
\caption{Input data model}
\end{table}  

Fields with plain-text details are of primary interest in this paper: (i) short description of an incident, (ii) detailed description of the incident, and (iii) reproduction steps.

The tool expects users to map fields to MS Excel columns. Also, users can configure name of the worksheet, date and time for data ingestion as shown below:
\begin{figure*}
	\centering
		\includegraphics[width=\textwidth]{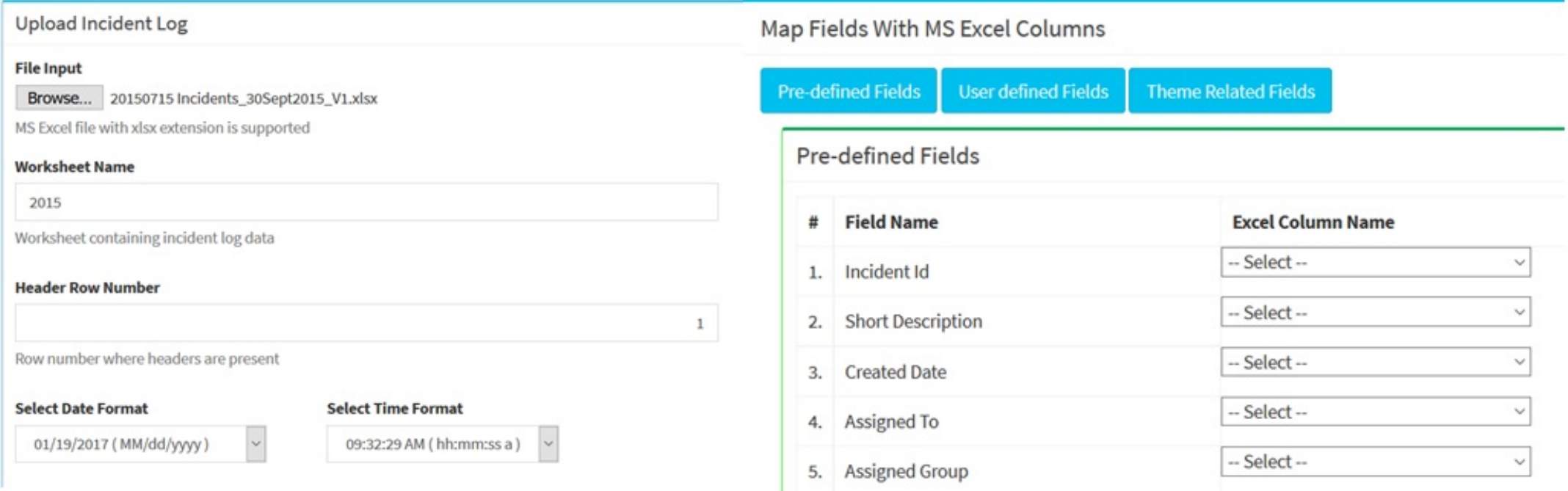}
		\caption{Data Ingestion Configuration}\label{settings}
\end{figure*}

Once the data is ingested then tool is ready for providing ranked list of similar tickets for a new bug, details of which are given to the tool as search query. Here user can apply filters if configured earlier and narrow down search by applying date range as shown below:
\begin{figure}
	\centering
		\includegraphics[scale=0.6]{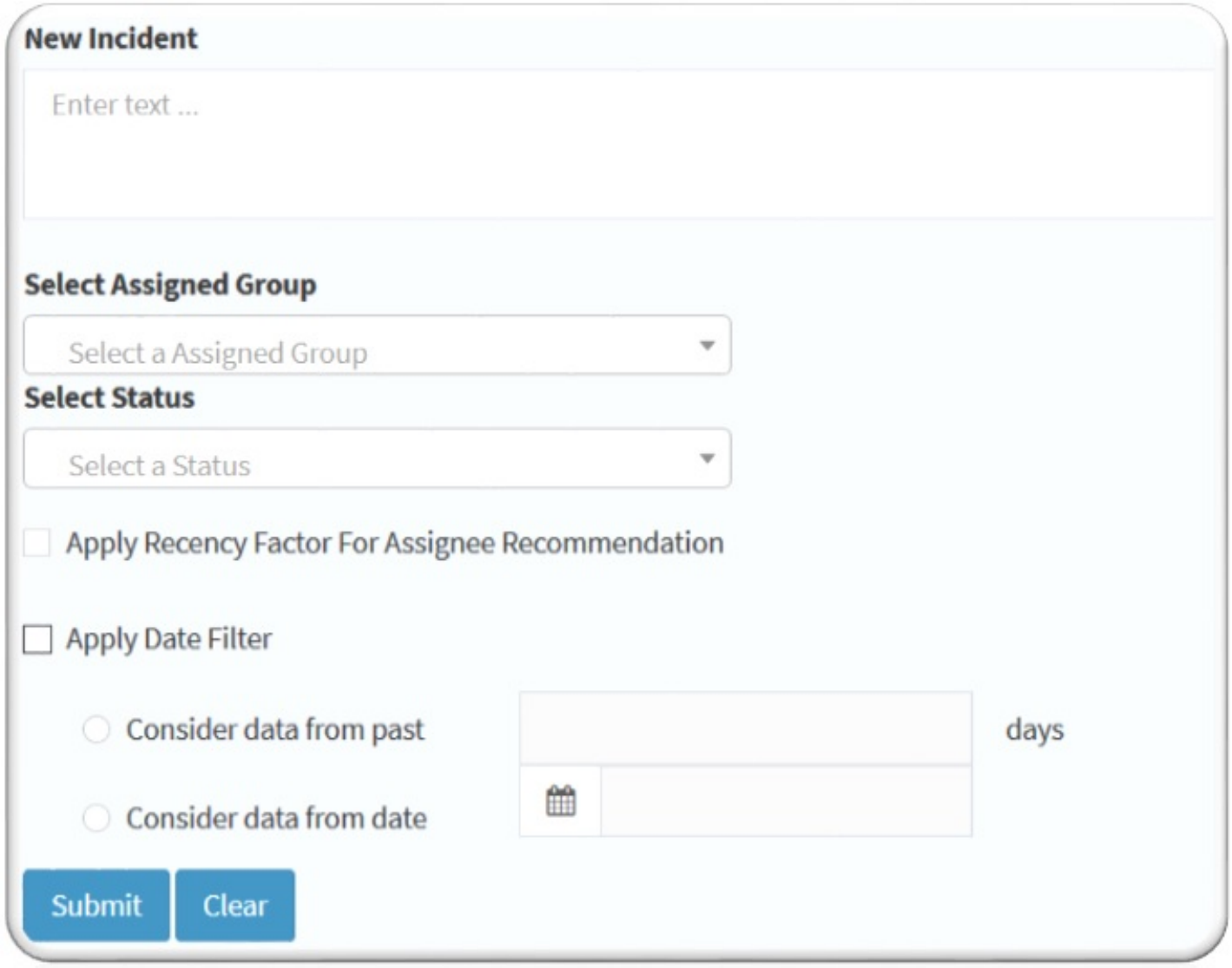}
		\caption{Search User Interface}\label{ingest_op}
\end{figure}

The results are categorized into four classes and users can get details of each output. The threshold of those four categories were decided based on empirical studies performed internally on various data sets.
\begin{figure*}
	\centering
		\includegraphics[width=\textwidth]{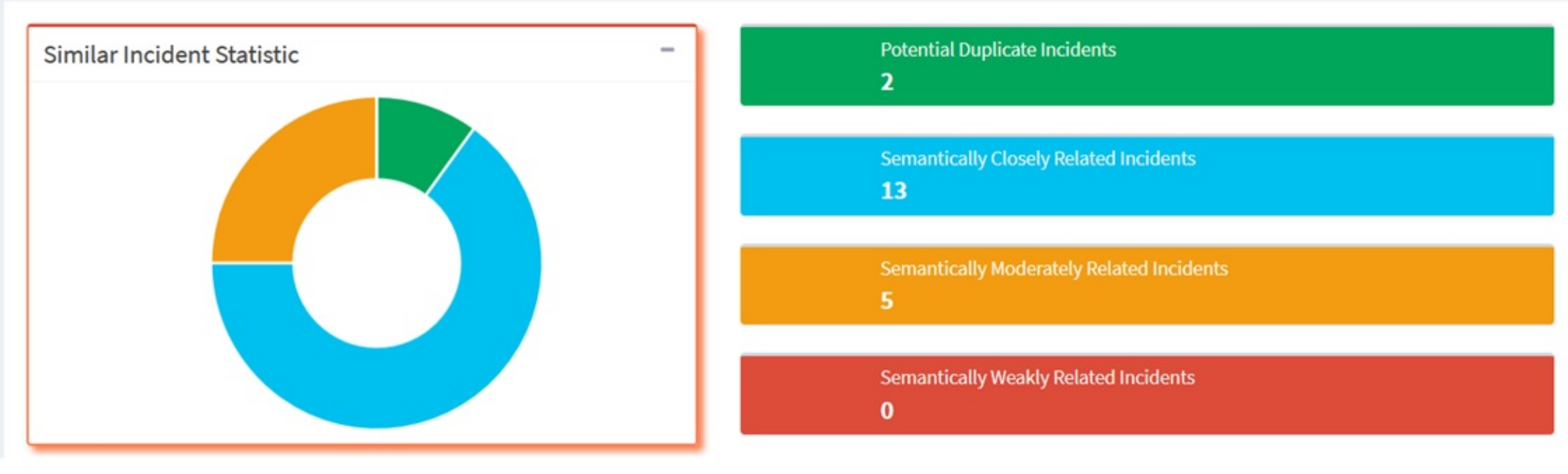}
		\caption{Search results}\label{results}
\end{figure*}

\section{Pilot Studies and Observations}\label{exp}

For experimental analysis, PoC tool was given to SMEs for evaluation in their actual practice from multiple client accounts spread across various industries including retail, leisure, energy and petrochemicals, and mining. In each case, we asked SMEs to consider multiple projects satisfying one or more of the following requirements: 
\begin{enumerate}
\item Rate at which bugs are being raised are not very low (< 10 per week) 
\item Historical bug data-base contains at least 100 tickets 
\item Degree of redirection of bugs is at least 10\% i.e., bugs are not getting closed correctly for at least 10\% cases 
\item There exist sizable number of bugs affecting multiple functionality and/or modules     
\end{enumerate}

Let us discuss for each use case eventual qualitative outcomes in light of the specific nature of the input data and other circumstantial factors which mattered in practice and challenges in adoption of current research in practice.

\subsection{Identifying Semantically Similar Bugs} 

\sf{\bf Nature of Data:} Most of the time only single line details were available as bug description apart from information on other fields (see Table 1). Detailed descriptions were not available in all cases or were not very different from brief descriptions. Resolution notes or root cause analysis or steps to reproduce bugs were also often missing or were very brief or very extensive involving detailed communication which took place (often using emails) among team members while resolving bugs. 

Additionally, plain text descriptions contained specific symbols (e.g. '\#* ... *\#') to enclose terms (meaning of which known only within team) as visual clues about the nature of bugs. Sometime bugs having exactly same text but differing in these localized terms were considered semantically related but not duplicate by SMEs. Also, during experimental analysis of many other use cases it turned out that such code words contained critical semantic information, however, this was something not possible to extract without SME help. 

\sf{\bf Experimental Analysis:} Table~\ref{t_dedup} presents results from a pilot study with a retail account, wherein tool was injested with ticket log of nearly twenty six thousand incidents and SMEs designed $59$ sample bugs as queries based upon recent tickets which were not part of the ingested ticket log and evaluated relevance of the outputs from the tool from view point of resolving these bugs. Since it was not known to SMEs how many relevant bugs actually existed in the ticket log, only precision of the outputs at two different levels (top 5 and top 20) was measured.
\begin{table*}[ht]
\centering
	\begin{tabular}{l|l|l}
\hline
\hline
& {\bf Relevant Results} & {\bf Relevant + Related Results} \\
\hline
\hline
\textbf{Top 5} & 67\% & 71\% \\
\hline
\textbf{Top 20} & 61\% & 69\% \\ \hline
\end{tabular}
	\caption{Precision of Search Results towards their relevance for users to identify duplicate bugs or identifying potential resolutions}
	\label{t_dedup}
\end{table*}

\sf{\bf Requirements for Minimal Solution:} In practice, domain familiarity and past experience of the users mattered a lot and often junior team members which were in sizable numbers, needed help from senior team members to figure out how to start understanding a bug. Therefore, when we asked them to use the tool, their primary requirement was that tool should be able to provide critical clues regarding a new bug, which otherwise would have come from senior SMEs.  

From operational perspective, users across levels appreciated that tool had small learning curve and one could start using it right away after installation with minimal configuration requirements. 

\sf{\bf Challenges in Adoption of Research:} When authors were designing the solution, it was not clear whether we need to adopt techniques which are claimed to be achieving best performance on open source data sets(~\cite{defectdatabse}) for the problem of duplicate defect identification. Interestingly most of the published works with high scores in detecting duplicate bugs apply supervised machine learning techniques (including deep learning), which though work well on large open source data sets, however, can't be easily adopted in practice without incurring significant cost to create desired training data in form of semantically equivalent pairs of bugs by the experienced SMEs before tool could be adopted in practice. Interestingly, parallel efforts were in place in some other client account towards adoption of such supervised ML based tools and in detailed comparative analysis, it turned out that our PoC tool was only 5\% below the other tool in performance (precision of top 20 relevant bugs for a new bug). On the other hand, simpler approaches based upon information retrieval methods have also been found to be reasonably effective in industrial settings~\cite{amoui2013search} and our experience confirms this.  

\subsection{Assignee Recommendation} 

\sf{\bf Nature of Data:} Most often ticket data only contains name or employee ID of the team member who raised and/or fixed a bug. However, when a team member is selected for an assignment, external factors like documented skills as recorded in HR databases, team-organization, current work load, etc. also play very important role in decision making process, which however are not recorded as a part of ticket details. 

\sf{\bf Requirements for Minimal Solution:} A solution which brings additional insights from the analysis of historical evidence received positive attention from the AM management. Time based filtering options were found to be of specific help since in many cases, at least some similar or related bugs had been fixed during recent past and older bugs might have been fixed by those who were no longer part of the team currently. 

\begin{figure}
	\centering
		\includegraphics[scale=0.6]{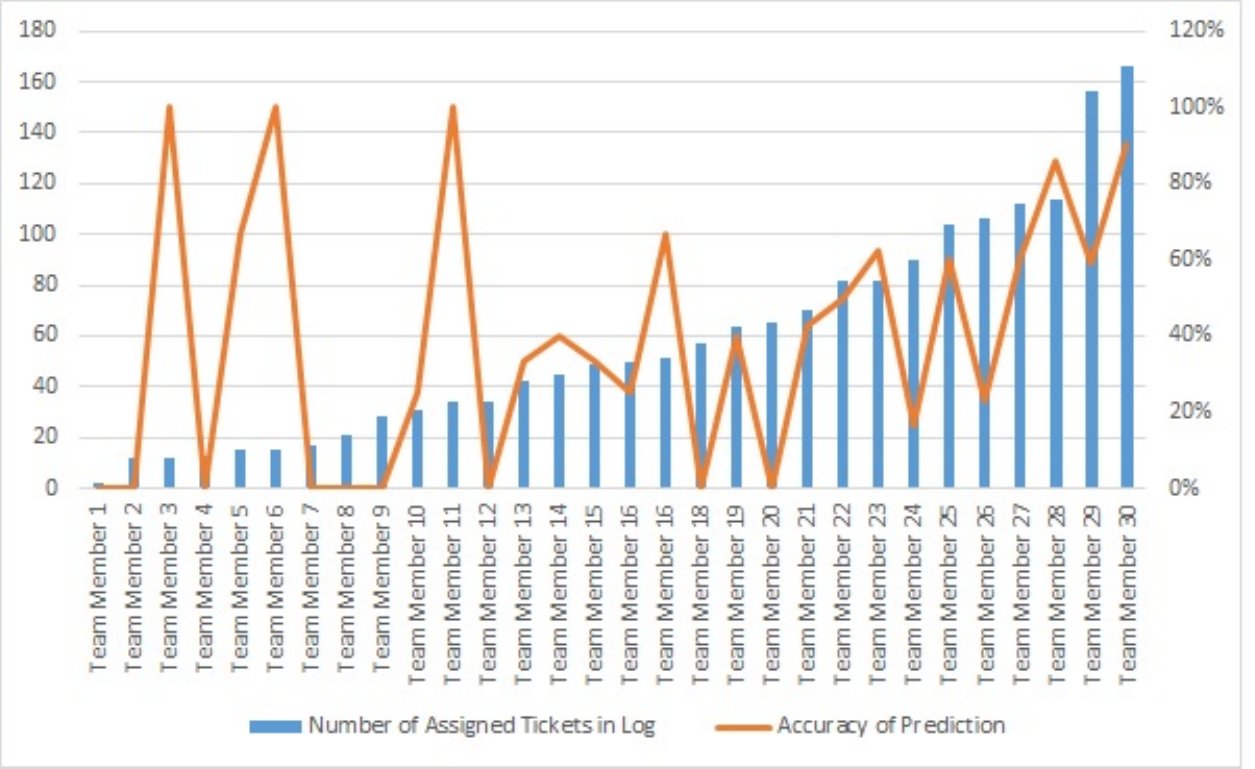}
		\caption{Graphical representation of association between tickets on which a team member has worked and accuracy in prediction of that team member during validation}\label{fig:assignee}
\end{figure}

\sf{\bf Experimental Analysis:} Table~\ref{t_assignee_1} presents results from pilot studies from three different industry accounts with varying ticket log sizes. In each case, we randomly identified more than 30 existing tickets from each log for validation purpose and estimated accuracy of predicting assignee when compared with the actual assignment as recorded in the log data. One challenge in this analysis is that our target set is relatively narrow (among many capable team members, only one is listed in the log-data set and we need to predict that!) Primary reason for adopting this intrinsic validation as against SME based validation was that SMEs expressed their lack of familiarity with the team members who worked long back and in some cases teams had just acquired the contract to maintain the application from another firm, hence SMEs did not have much knowledge to validate outputs.   
\begin{table}[ht]
\centering
	\begin{tabular}{l|l|l|l}
\hline
\hline
{\bf Data Set} & {\bf Size of} & {Number of} & {\bf Accuracy} \\
 & {\bf Ticket Log} & {Assignees} & \\
\hline
\hline
\textbf{Data-Set 1} & 1850 & 31 & 0.50 \\ \hline
\textbf{Data-Set 2} & 10670 & 13 & 0.47 \\ \hline
\textbf{Data-Set 3} & 166902 & 26 & 0.47 \\ \hline
\end{tabular}
	\caption{Average accuracy in predicting assignee using historical log as reference data}
	\label{t_assignee_1}
\end{table}
Realizing that on average results were relatively poor, we investigated this further to learn whether there existed any correlation between number of tickets which were assigned to a team member and accuracy in being able to predict that team member as recommended assignee during validation. Graph~\ref{fig:assignee} depicts that visually and as a numerical indicator, Pearson correlation coefficient between these factors was estimated as $0.34$, which clearly indicated that higher number of assigned tickets to a team member may not imply higher accuracy in predicting whether a team member should work on a ticket. This could be owing to the factors which were not recorded in the ticket log, however were playing key role during ticket assignment.   
\sf{\bf Challenges in Adoption of Research:} Despite the fact that there exist many published research works on this problem, in practice today, often ad-hoc approaches reflecting prevailing organizational processes and practices are to be found. One reason for this is that often there are some transient and circumstantial factors, which affect the decision making process. Hoever, such factors are not recorded in a way that these could be given as inputs to a tool. For example, consider the role of training programs, which play a role while assigning tickets to those team members who had undergone training on related functional areas recently. However, it would be hardly captured in an AM tool. This became further evident when we tried training a ML model on the ticket log (using features from vector space modeling and TF-IDf based information theoretic weighing) and Table~\ref{t_assignee_2} presents the results. It is clear from these results that ticket details as captured in the AM tool are hardly sufficient for achieving high accuracy in automated assignee recommendation. However, in practice, these outputs can still add value in assisting users in making informed decisions. 
\begin{table}[ht]
\centering
	\begin{tabular}{l|l|l|l|l}
\hline
\hline
{\bf Size of} & {\bf Number of} & {\bf P}& {\bf R} & {\bf F1-Score}\\
{\bf Ticket Log} & {\bf Assignees} & &  & \\
\hline
\hline
5600 & 28 & 0.35 & 0.31 & 0.33 \\ \hline
\end{tabular}
	\caption{10 fold cross validation of SVM Classifier for Assignee Prediction}
	\label{t_assignee_2}
\end{table}

\subsection{Theme Mining} 

\sf{\bf Nature of Data:} Ticket log data often contains primary classification of the bugs in terms of modules or types of bugs. However, it misses giving any further insights, which can only be gleaned by detailed analysis of the bug descriptions, priority levels, number of re-directions, priority-escalations etc. 

\sf{\bf Requirements for Minimal Solution:} Themes or central problem areas must be extracted such that their correlations with user categorizations are explicit. Furthermore, tool must provide detailed navigation support so that users get explanatory support underlying extracted results by the tool. For example, if a tool depicts that a problem area {\it p = `page access issue'} is weakly connected with problem area {\it q = `password reset'} but is strongly connected with {\it r = `login failure'}, user should be given underlying justification in terms of displaying list of tickets where first pair of problem areas (p, q) appear together versus where second pair (p, r) of problem areas appear in the ticket log.

{\sf Experimental Analysis:} Table~\ref{t_tm} shows results of various techniques, which we experimented with for mining central problem areas using examples ticket log consisting of twenty six thousand ticket details. Considered techniques included including latent semantic analysis based centrality; latent Dirichlet allocation; term frequency, and term frequency  and inverse document frequency based centrality. Themes were extracted using textual fields as source of analysis. These themes were mined and analyzed for whole ticket-log and compared with project SMEs specified tagging. Precision of mined themes was not considered since SMEs were not available for analysis of the results and it would have been incorrect to assume that any theme not present in user specified tags must be considered wrong. 
\begin{table}[ht]
\centering
	\begin{tabular}{l|l}
\hline
\hline
{\bf Approach} & {\bf Recall}\\
\hline
\hline
LSA + TF & 48\% \\ \hline
LSA + TF\_IDF &	48\%  \\ \hline
LSA + LDA (Mallet)	& 48\%  \\ \hline
LDA + TF	& 43\%  \\ \hline
LDA + TF\_IDF	& 43\%  \\ \hline
TF &	38\%  \\ \hline
TF\_IDF & 38\%  \\ \hline
LSI & 38\%  \\ \hline
LDA & 33\%  \\ \hline
\end{tabular}
	\caption{Recall Levels for Theme Mining approaches (top 50 results compared to user specified 21 themes. LSA: latent semantic analysis; LDA: latent Dirichlet allocation; TF: Term Frequency; TF\_IDF: Term Frequency Inverse Document Frequency)}
	\label{t_tm}
\end{table}
  
\sf{\bf Challenges in Adoption of Research:} As can be seen from relatively poor results in Table~\ref{t_tm}, primary challenge in adoption of existing approaches (mostly unsupervised in nature) is that they do not appear to align well with users' expectations. Alternative approaches which aim at filling this gap are supervised ML based approaches, which however require significant amount of annotated data to start with.  

\subsection{Business Process Mapping} 

\sf{\bf Nature of Data:} In practice, it turned out that often mapping of bugs with standard business processes is not mentioned and only proxy mappings with modules or components are recorded in AM tools. However, in case of large scale applications, it can be assumed that component or module level architecture approximately aligns with underlying business process hierarchy and relationships. Furthermore, assuming that business process maps are generally represented as a tree (acyclic undirected graph), it is sufficient to consider leaf level business process nodes (i.e., business process steps) alone as representative for who business process flow-path. 

{\sf Experimental Analysis:} Table~\ref{t_bpm} shows results of ML based prediction approach for associating business processes for tickets based upon learning from historical log data. We employed 10-fold cross validation as evaluation measure and experimented with six different data sets from different industry accounts with sizes of ticket logs ranging from 448 to 22000. 

\begin{table}[ht]
\centering
	\begin{tabular}{l|l|l|l|l|l}
\hline
\hline
\textbf{Ticket-Log} & \textbf{\#BP}	& \textbf{\#Tkts} & \textbf{P} & \textbf{R} & {\bf F1-score} \\
\hline
\hline
Data-Set\#1 & 14	& 448		& 0.67	& 0.66 & 0.66  \\ \hline
Data-Set\#2 & 11	& 22000	& 0.78	& 0.76 & 0.77  \\ \hline
Data-Set\#3 & 3	& 6420	& 0.86	& 0.85 & 0.85    \\ \hline
Data-Set\#4 & 8	& 3600	& 0.67	& 0.67 & 0.67    \\ \hline
Data-Set\#5 & 5	& 5630	& 0.54	& 0.53 & 0.53    \\ \hline
Data-Set\#6 & 4	& 2000	& 0.90	& 0.90 & 0.90    \\ \hline
\end{tabular}
	\caption{Precision, Recall, and F1-Scores for SVM (Gaussian kernel) classifier based ticket to business process mapping problem. \#BP: No. of Business Processes; \#Tkts: Number of tickets}
	\label{t_bpm}
\end{table}

Noticing that performance of the tool was comparatively poorer in Data-Set\#5, we investigated the cases where tool predicted incorrectly and it turned out that in most of the cases, bug description, which formed primary feature space for the ML technique, consisted of cryptic terms (often unique) like `Kfeil: JSP045ABCD \#ABCDD11759 [(231630 07/32/15), (0AAAAAAAAAAXXZW2M2U)]. ABCD1168 (\#J137891)'. We extended this analysis to other Data-Sets as well and realized that presence of such terms, which can't be parsed without SME intervention, was primary source of errors in outputs.  

\sf{\bf Challenges in Adoption of Research:} Application of supervised ML based approach resulted into comparatively better results and could be adopted in practice if designed as a separate tool which works in conjunction with existing AM tools. However, close integration with AM tools is not always possible since not all proprietary tools expose APIs for automated data export and in those cases, manual efforts in exporting data from AM tool and feeding it into tools like CogAM is necessary.  

\subsection{Threats to Validity} Even though we tried working with multiple industry accounts and had variety of data to analyze, it might still be limited by way of being influenced by circumstantial organizational factors which determined the nature of projects which we could work with currently. Furthermore, nature of software applications' data is a highly dynamic in nature and hence temporal reliability of the results would definitely demand ongoing reevaluations.      

\section{Conclusion and Future Work}\label{con}
In this paper, we discussed experiences with a data-driven application maintenance tool called CogAM, which applies information retrieval based semantic search technology and unsupervised / supervised machine learning approaches for addressing spectrum of interrelated problems in AM domain. In specific, we tackle the critical problems of checking duplicate incidents, assignment of tickets to appropriate resources, mining the central themes for a cluster of incidents, and mapping incidents to the underlying business processes. 

Today, each of these tasks are carried out manually with considerable inefficiency and cost. While some of these problems have been researched extensively with sophisticated techniques on open source repositories and boards like Apache Foundation or Stack Overflow; we hardly find them to be shop usable as they impose challenges for adoption. Approach and tool has been piloted in real customer projects and we have received valuable lessons in terms of fitment, usability and adoption. Our experience suggests that in some cases search based solutions are good enough while in others detailed machine learned classifiers with annotated data would provide better results. Still in others, the problem becomes difficult as the ticket data as a single source of information is not enough to perform the necessary tasks which depend on external circumstantial inputs. In all cases, a richer domain knowledge becomes a key ingredient to success.  

In future, we propose to carry out research in two main directions. First, there is clear need to carry out extensive experimentation with mass deployments to study the problem patterns across different types of maintenance projects and gather further insights. Second, we realize that rather than using pre-trained models, it would be better to experiment with active learning approaches where the engineers can interact and provide iterative supervision for better contextual refinement.   

\bibliographystyle{abbrv}
\bibliography{TIA_RC} 

\end{document}